\documentclass[a4paper,10pt]{article}
\usepackage{graphicx}
\usepackage[left=2.2cm, right=2.2cm, top=3cm]{geometry}
\pdfoutput=1

\title{Single-layer Compton detectors for measurement of polarization correlations of annihilation quanta\footnote{Accepted for publication in Nuclear Instruments and Methods in Physics Research Section A: Accelerators, Spectrometers, Detectors and Associated Equipment, https://doi.org/10.1016/j.nima.2019.162835 \newline {\textsuperscript\textcopyright}2019. This manuscript version is made available under the CC-BY-NC-ND 4.0 license http://creativecommons.org/licenses/by-nc-nd/4.0/ }} % Article title

\author{Mihael Makek\textsuperscript{1}\footnote{Corresponding author: makek@phy.hr}, Damir Bosnar\textsuperscript{1}, Luka Paveli\'{c}\textsuperscript{1,2}, \\
Pavla \v{S}enjug\textsuperscript{1}, Petar \v{Z}ugec\textsuperscript{1}} % Authors
\date{}

\begin{document}

\maketitle % Print the title and abstract box

\begin{flushleft}
\textsuperscript{1}\textit{Department of Physics, Faculty of Science, University of Zagreb, Bijeni\v{c}ka c. 32, 10000 Zagreb, Croatia} \\ % Author affiliation 
\textsuperscript{2}\textit{Institute for Medical Research and Occupational Health, Ksaverska cesta 2, 10000 Zagreb, Croatia}  \\ % Author affiliation
\end{flushleft}

%----------------------------------------------------------------------------------------
%	ARTICLE CONTENTS
%----------------------------------------------------------------------------------------

\begin{abstract}
Measurement of gamma-ray polarization can provide valuable insight in different areas of physics research. One possible application is in Positron Emission Tomography, where the annihilation quanta with orthogonal polarizations are emitted.  Since polarization can be measured via Compton scattering, the initial orthogonality of polarizations can be translated to correlation of azimuthal scattering angles, and this correlation may be exploited as an additional handle to identify the true coincidence events. In order to examine the concept of utilizing the polarization correlations in PET, we have used a system of two compact, position and energy-sensitive Compton scattering detectors in coincidence mode. Each consists of a single matrix of scintillation pixels, read-out by a matching array of Silicon photomultipliers on the back side. The Compton events in each module are clearly identified and the scattering angles are reconstructed from the energy deposition and event topology. We have extracted the polarimetric modulation factors from the measured azimuthal scattering angles of the two Compton-scattered gammas and studied their dependence on Compton scattering angles $\theta$ and on azimuthal resolution $\Delta\phi$. For scattering angles around $\theta_{1,2}=82^\circ$, where the maximum modulation is expected, the modulation factors from $\mu=0.15\pm 0.01$ to $\mu=0.27\pm 0.02$ have been measured, depending on the azimuthal resolution, which is governed by event topology in the detectors. Analogously, for scattering around $\theta_{1,2}=70^\circ$, modulation factors from $\mu=0.12\pm 0.01$ to $\mu=0.21\pm 0.02$ have been obtained. The results show that the measurement of the polarization correlations of annihilation quanta are feasible with compact single-layer, single-side read-out detectors, which may be used to build cost-efficient systems for various applications where gamma-ray polarization information is of interest.
\end{abstract}


\section{Introduction}
Gamma-ray polarization measurement relies on Compton scattering, where according to Klein-Nishina cross-section, the most probable azimuthal scattering angle of the gamma is perpendicular to the incident polarization vector. Although explored for astrophysics (e.g. \cite{lei1997, mitani2004, spillmann2008, bloser2009, takeda2010, yonetoku2011}), the polarization of gammas has not been implemented in biomedical imaging. A potential use can be Positron Emission Tomography (PET), where gammas emitted from $e^+e^-$ annihilation have initially orthogonal polarizations. If both undergo Compton scattering, the orthogonality of their polarizations will result, with a high probability, in orthogonality of their azimuthal scattering angles. Since the polarization is independent of energy, this azimuthal (polarization) correlation offers another independent handle to identify the true coincident events. Preliminary studies have shown exploiting this feature has a potential to contribute to the image quality of a PET system, especially with sources of high activity where the probability of a false positive coincidence is significant \cite{kuncic2011}. A Monte-Carlo model of a PET system utilizing polarization correlations has been developed to show the feasibility of the approach \cite{mcnamara2014, toghyani2016}. To date, however, this has not been demonstrated experimentally.

In this paper we present that measurement of polarization correlations of annihilation quanta is possible with detector modules encompassing a single array of scintillation pixels and silicon photomultipliers (SiPM). The single-layer concept makes the modules compact and cost-efficient compared to more common dual-layer Compton detectors, and opens possibility for their use in applications where gamma-ray polarization information is of interest. 

\section{Measurement of Polarization Correlations}
Two photons originating from $e^+e^-$ annihilation are emitted back-to-back with 511 keV energy and orthogonal polarizations. In case both of them undergo Compton scattering with scattering angles $\theta_{1,2}$ and azimuthal angles $\phi_{1,2}$, respectively, the differential cross-section is given by \cite{pryce1947, snyder1948}:
\begin{equation}
 \frac{\mathrm{d}^2\sigma}{\mathrm{d}\Omega_1\! \mathrm{d}\Omega_2}\!=\!\frac{r_0^4}{16} F(\theta_1\!)F(\theta_2\!) \!\left\lbrace \! 1\! - \! \frac{G(\theta_1\!)G(\theta_2\!)\!}{F(\theta_1\!)F(\theta_2\!)} \cos[2(\!\phi_1\!-\!\phi_2\!)] \right\rbrace
 \label{eqn:K-N}
\end{equation}
with 
\begin{eqnarray}
 F(\theta_i)=\frac{[2+(1-\cos\theta_i)^3]}{(2-\cos\theta_i)^3}, \;\;
 G(\theta_i)=\frac{\sin^2\theta_i}{(2-\cos\theta_i)^2}
\end{eqnarray}
where $i=1,2$. Since initially, the polarization vectors of both photons are orthogonal, the cross-section has the maximum for $|\phi_1-\phi_2|=90^\circ$ (for the fixed scattering angles $\theta_{1,2}$), hence the polarization correlation is preserved in the Compton scattering process. The sensitivity of the measurement to the initial polarizations is characterized by the polarimetric modulation factor, defined as:
\begin{equation}
 \mu \equiv \frac{P(\phi_1-\phi_2=90^\circ) - P(\phi_1-\phi_2=0^\circ)}{P(\phi_1-\phi_2=90^\circ) + P(\phi_1-\phi_2=0^\circ)}
\end{equation}
where $P(\phi_1-\phi_2=90^\circ)$ and $P(\phi_1-\phi_2=0^\circ)$ are the probabilities to observe the two scattered gammas with perpendicular and parallel azimuthal angles, respectively. 

The modulation, $\mu$, depends on Compton scattering angles $\theta_{1,2}$ and it reaches the maximum $\mu_{max}=0.48$ for $\theta_1=\theta_2 \approx 82^\circ$ \cite{snyder1948}. Although the correlation is the strongest at $\theta_1=\theta_2 \approx 82^\circ$, one has to take into account the probabilities to have Compton scattering at those angles (i.e. the cross-section) in order to take the optimal advantage of the polarization correlation as a tool to recognize the true coincident events in PET. For the 511 keV gamma photons, the cross-section for single photon Compton scattering peaks at forward angles, $\theta\approx34^\circ$. 
Therefore, the region around $\theta_{1,2}\approx70^\circ$ has been suggested as optimal \cite{toghyani2016}, since it provides a relatively high scattering probability and a relatively strong polarimetric modulation factor, $\mu=0.40$. It has to be noted that the modulation factors are somewhat reduced in finite (realistic) detector geometries, since they are integrated over the acceptance of the detectors \cite{snyder1948}.

Experimentally, we can determine the polarimetric modulation factor by measuring the distribution of azimuthal angle difference, $N(\phi_1-\phi_2)$, for given range of scattering angles $\theta_1$ and $\theta_2$. The observed distribution must be corrected for the non-uniformities in detector acceptance, according to:
\begin{equation}
 N_{cor}(\phi_1-\phi_2) = \frac{N(\phi_1-\phi_2)}{A_n(\phi_1-\phi_2)}
 \label{eqn:Nnorm}
 \end{equation}
where the $A_n(\phi_1-\phi_2)$ is the normalized acceptance for the particular azimuthal angle difference, which can be obtained from simulation or experimentally by measuring the distributions of randomly-polarized sources. We adopt the latter approach, where $A_n(\phi_1-\phi_2)$ is obtained by event-mixing technique in which the azimuthal difference $(\phi_1-\phi_2)(mixed)$ is reconstructed using $\phi_1$ and $\phi_2$ from different events. The orientation of the polarization of single annihilation quantum does not have a preferred direction, so different events will have gammas with randomly oriented polarizations. Therefore, $A_n(\phi_1-\phi_2) (mixed)$, does not contain the polarization correlation of annihilation quanta, but keeps the information of the detector pair acceptance, and can be used as the acceptance correction.  

It has been shown in \cite{toghyani2016} that $\mu$ equals the $\frac{G(\theta_1) G(\theta_2)}{F(\theta_1) F(\theta_2)}$ from Eqn. \ref{eqn:K-N}. Hence, according to equation \ref{eqn:K-N}, experimentally one expects:
\begin{equation}
 N_{cor}(\phi_1-\phi_2) = M [1 - \mu \cos (2\phi_1-\phi_2)]
\label{eqn:Ncor} 
\end{equation}
where $M$ will correspond to the average amplitude of the distribution and $\mu$ to the modulation factor.

\section{Experimental Setup}
\begin{figure}
\centering
\includegraphics[width=8.4cm]{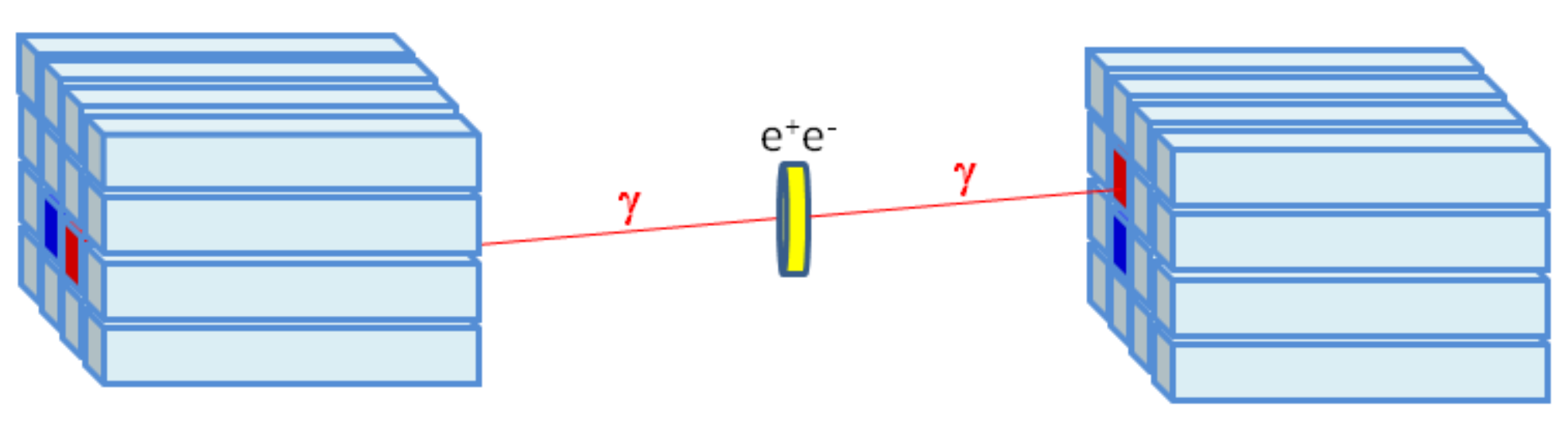}
\caption{Visualization of two scintillation pixel modules (not to scale).}
\label{fig:visual2}
\end{figure}

We have set up a system of two modules (Figure \ref{fig:visual2}), each encompassing a 4x4 matrix of Lutetium Fine Silicate (LFS) scintillators (pixel size 3.14 x 3.14 x 20 mm$^3$), with a 3.2 mm pitch. The system was tested with 511 keV gammas from $^{22}$Na-source. A matrix is read out by a SiPM array in a one-to-one match, where each channel is amplified and sampled at 1.6 GS/s. Under typical operating voltage of $U_b=U_{breakdown}+1.6 \;\mathrm{V}$ and temperature $t=20^\circ\!-\!22^\circ \; \mathrm{C}$, the modules achieve an average energy resolution of $\Delta E=12.2\% \pm 0.7\%$ (FWHM) at 511 keV and the coincidence time resolution between the modules is $\Delta t=0.54 \pm 0.02$ ns (FWHM) for the annihilation gammas. The system components are described in more detail in \cite{makek2017}, while its performance is reported in \cite{makek2019}.

\begin{figure}
\centering
\includegraphics[width=8.6cm]{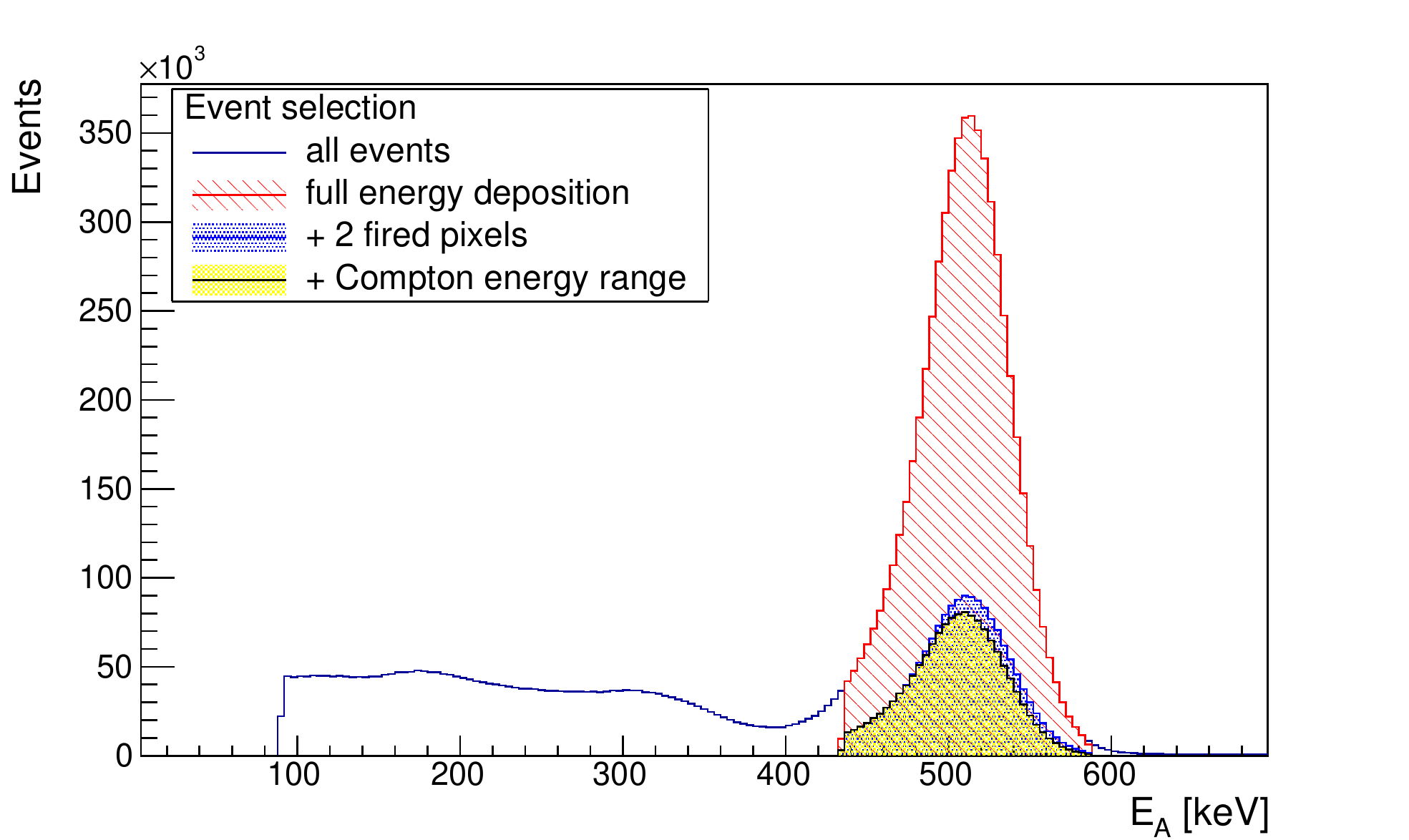}
\caption{Energy deposition of 511 keV gamma in module A, with various selection criteria.}
\label{fig:EA}
\end{figure}

\begin{figure}
\centering
\includegraphics[width=8.6cm]{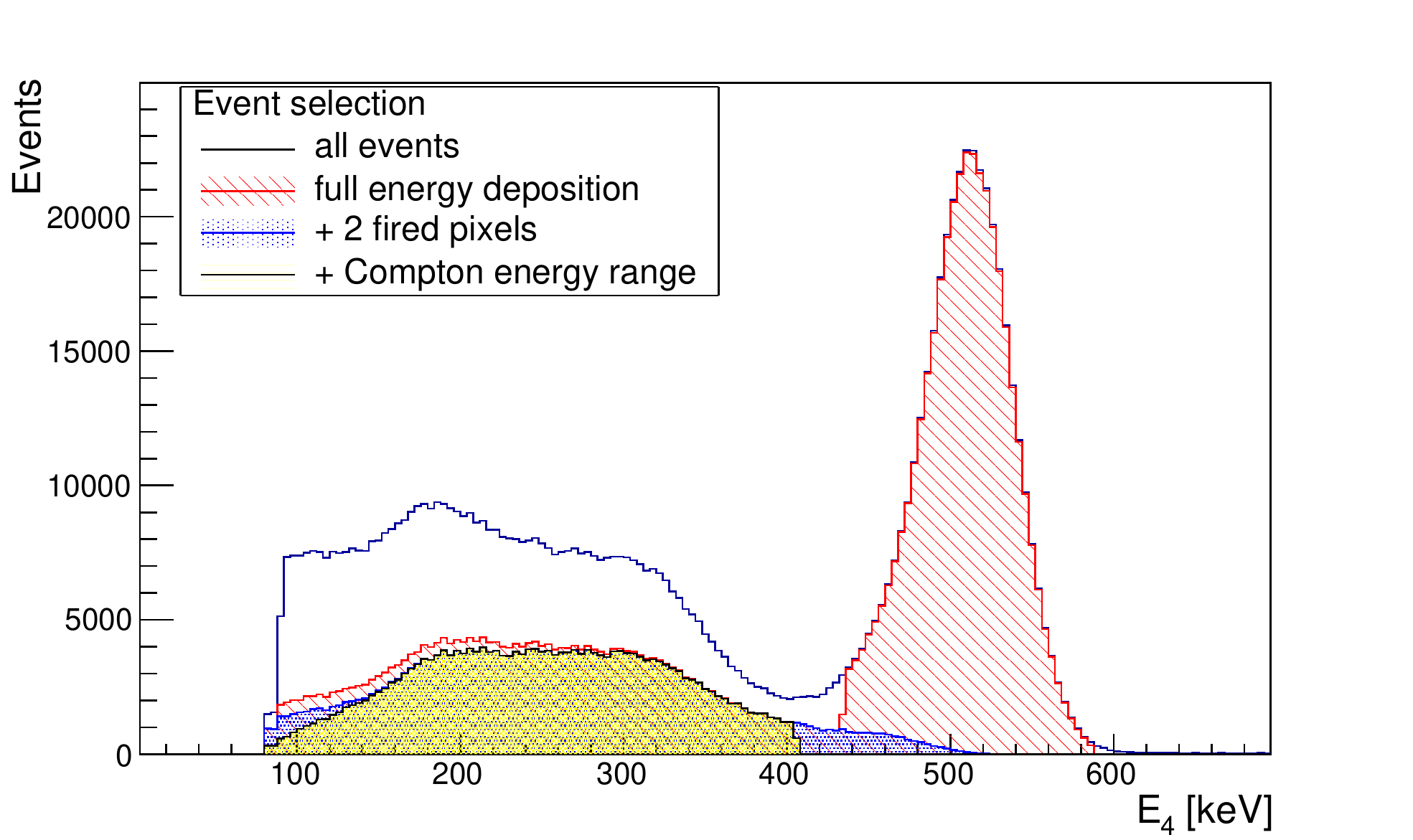}
\caption{Energy deposition of 511 keV gamma in a single pixel (no. 4), with various selection criteria.}
\label{fig:E4}
\end{figure}

\begin{figure}
\centering
\includegraphics[width=8.8cm]{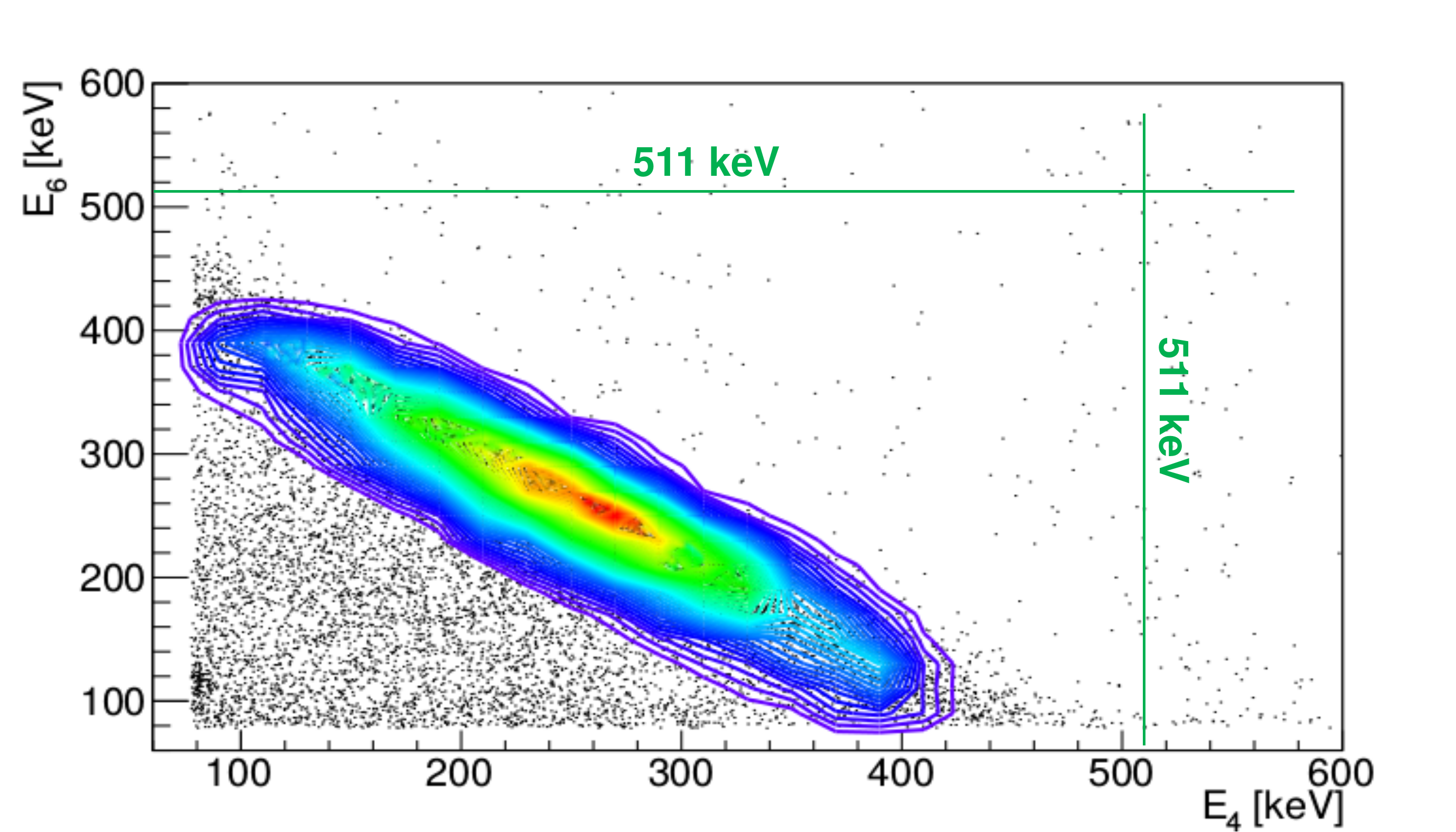}
\caption{Energy deposition of 511 keV gamma in pixel 6 vs. pixel 4: events above the threshold in both pixels (scatter), selected Compton events (contour).}
\label{fig:E4_E6}
\end{figure}

To select the Compton events, we require that the full gamma energy is deposited in a module (within $\pm3\sigma$) and that exactly two pixels fire with  energy $80\;\mathrm{keV}\! < E_{px}\! < 405\;\mathrm{keV}$, where the lower limit is set conservatively to reject noise and the upper limit is determined by the Compton edge (smeared by the energy resolution). The effects of these criteria on the spectra in a module and in a single pixel are shown in Figures \ref{fig:EA} and \ref{fig:E4}, respectively. The relation of energies deposited in two contributing pixels is shown in Figure \ref{fig:E4_E6}. 

In Compton events, the scattering angle, $\theta$ in each module, is reconstructed from the energies deposited in the pixels. For 511 keV gammas, if deposited energies are $E_{px}\! <\! 170\;\mathrm{keV}$ or $E_{px}\! >\! 341\;\mathrm{keV}$, the former should correspond to recoil electron (scattering) pixel and the latter to the scattered gamma (absorption) pixel, according to Compton scattering kinematics. In events with $170 \;\mathrm{keV}\! <\! E_{px}\! <\! 341\;\mathrm{keV}$, there may be an ambiguity in identification of the scattering and absorption pixel, since a scattering at a forward angle can result in the same energy responses of the pixels as a scattering at a backward angle. It was shown that this detector geometry and  material result in a higher probability to detect the forward scattering than the backward one, owing to the higher cross-section and the lower absorption probability of the scattered gamma in the former case\cite{makek2019}. Hence we always assume the forward scenario, assigning the pixel with the lower energy to the recoil electron and the one with the higher energy to the scattered gamma. The angular resolution is $\Delta \theta \simeq 18.8^\circ$ (FWHM) throughout the acceptance \cite{makek2019}. The azimuthal angle, $\phi$, is reconstructed from the relative position of the two fired pixels in a module. Its resolution is governed by the uncertainty of the interaction position within a pixel and the distance, d, of the fired pixels: $\sigma_\phi = \frac{1}{\sqrt{6}}\left|\frac{a}{d}\right|$, where $a=3.14$ mm is the pixel width \cite{makek2019}. The azimuthal resolution ranges from $\Delta \phi=54^\circ$ (FWHM) for the closest neighbors to $\Delta \phi=12.7^\circ$ (FWHM) for the most distant pixels. 

The measurement of polarization correlations was conducted with a $^{22}$Na-source ($\approx 1 \mu Ci$) enclosed in an aluminum case, placed on the central system axis, 4 cm from the front face of each module. The trigger accepted only the events where coincidence between the modules occurred. The presented analysis is based on 50 million recorded events, of which 19.4 million had full energy deposition in both modules and 1.05 million had passed the additional Compton event selection.  

\section{Results and discussion}
For the events where Compton scattering occurs in both detector modules, the scattering angles $\theta_{1,2}$ and azimuthal angles $\phi_{1,2}$ are reconstructed and the acceptance-corrected distribution of azimuthal angle difference, $N_{cor}(\phi_1-\phi_2)$ is obtained for the selected range of $\theta_{1,2}$. 

\begin{figure}
\centering
\includegraphics[height=4cm]{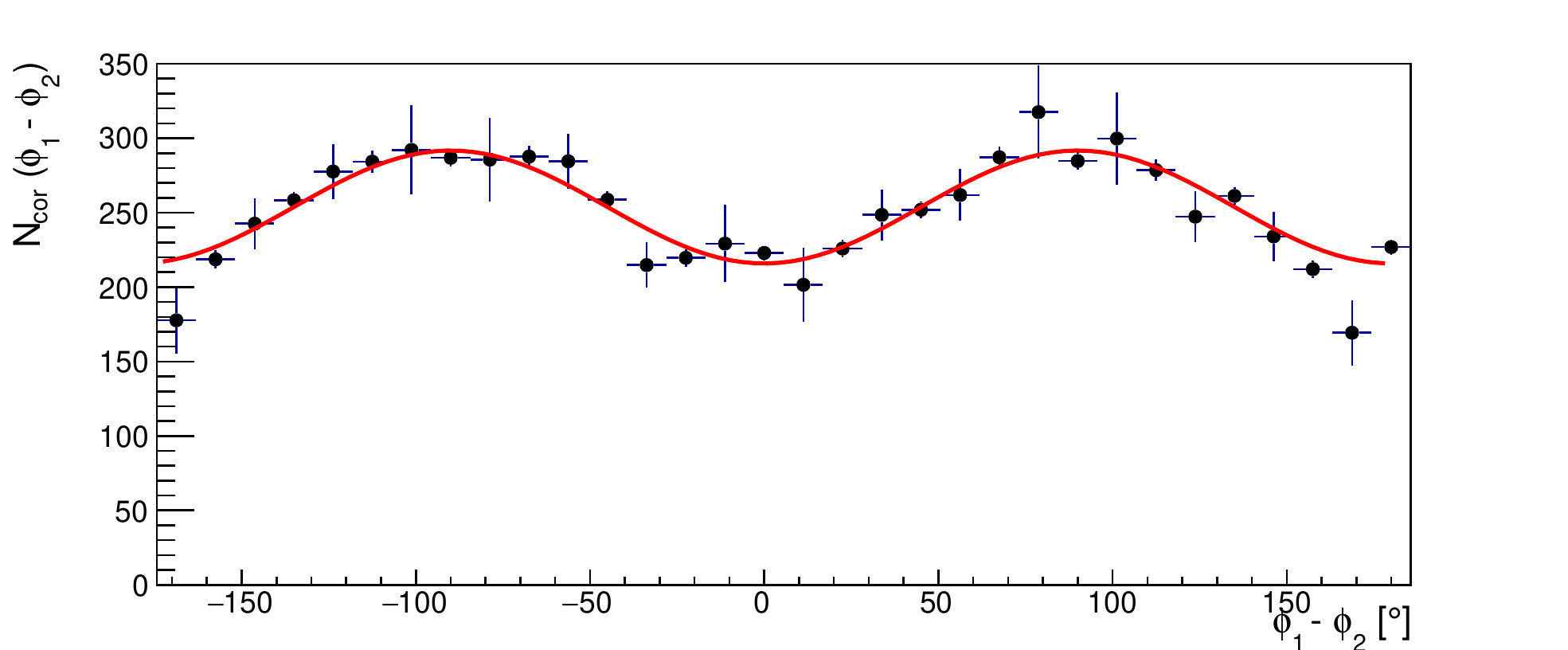}
\caption{The acceptance corrected $\phi_1-\phi_2$ distribution for $72^\circ\! \leq\! \theta_{1,2}\! \leq\! 90^\circ$ and all pixel distances $3.2\;\mathrm{mm} \leq d \leq 13.6 \;\mathrm{mm}$. The line is a fit of Eqn. \ref{eqn:Ncor}.}
\label{fig:82_all}
\end{figure}

\begin{figure}
\centering
\includegraphics[height=4cm]{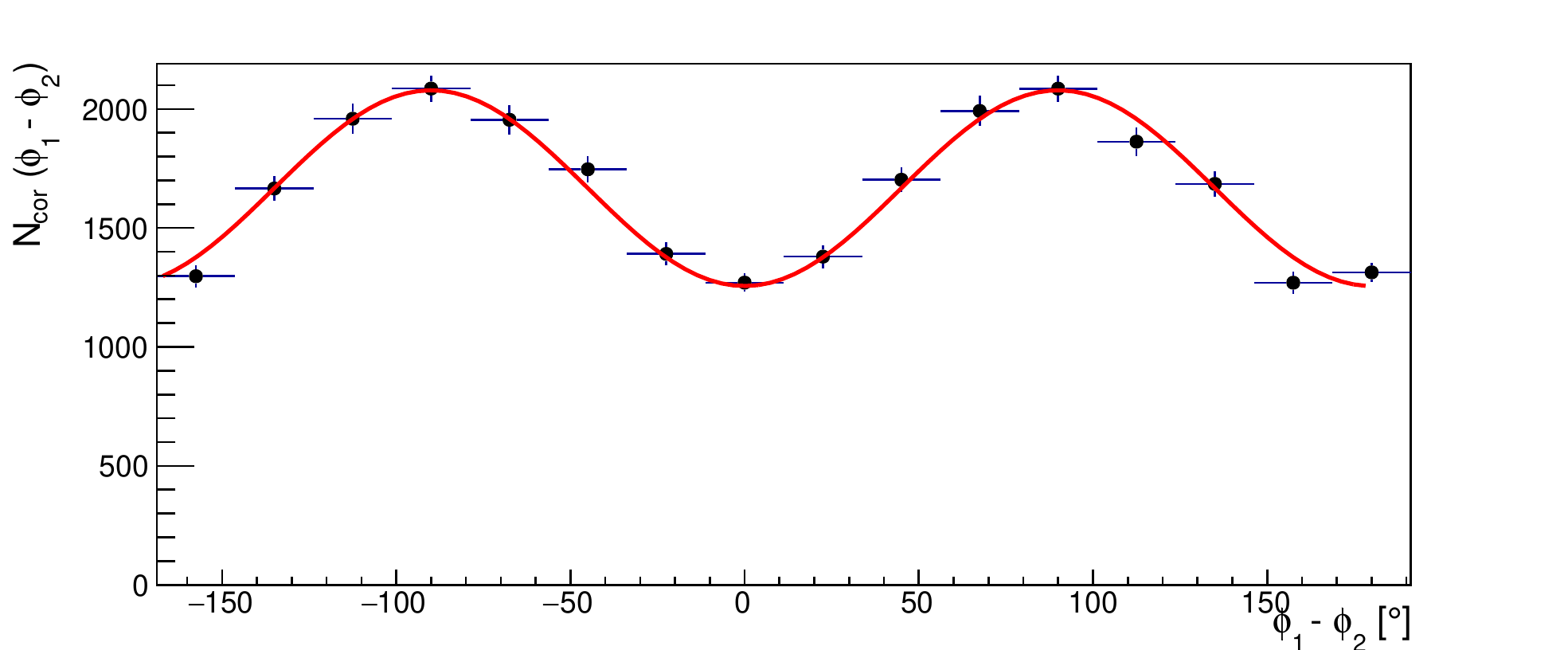}
\caption{The acceptance corrected $\phi_1-\phi_2$ distribution for $72^\circ\! \leq\! \theta_{1,2}\! \leq\! 90^\circ$ and for pixel distances $4.5\;\mathrm{mm} \leq d \leq 13.6 \;\mathrm{mm}$. The line is a fit of Eqn. \ref{eqn:Ncor}.}
\label{fig:82_near}
\end{figure}

First, we selected the scattering angles $72^\circ\! \leq\! \theta_{1,2} \!\leq\! 90^\circ$, centered around $\theta=82^\circ$ where the maximum azimuthal correlation is expected. The reconstructed distribution for all event topologies, corresponding to pixel distances $3.2\;\mathrm{mm} \leq d \leq 13.6 \;\mathrm{mm}$, is shown in Figure \ref{fig:82_all}. The error bars represent the contribution of the statistical and the systematic error added in quadrature. The latter is determined by examining the acceptance corrected yield at $-90^\circ,0^\circ,90^\circ, 180^\circ,$ in dependence of the histogram bin width and it is estimated to be 2\% of the yield. The distribution is fit with the function from Eqn. \ref{eqn:Ncor}, from which the modulation factor $\mu=0.15 \pm 0.01$ is obtained. Further, we explored the dependence of the modulation factor on event topology, i.e. on the distance of fired pixels, which determines the azimuthal resolution. The modulation factors obtained when pixels with specific distances are selected, are shown in Table \ref{tab:mu_82}, Set 1-4. It clearly shows that the lowest modulation factors are obtained for $d=3.2$, when the fired pixels are the adjacent neighbors, in which case the azimuthal uncertainty is the largest. If events with fired adjacent neighbors are not used, the measured modulation is significantly higher (Table \ref{tab:mu_82}, Set 6,7), as shown for example in Figure \ref{fig:82_near}. 

\begin{table}
 \centering
 \begin{tabular}{c | c c c} \hline
  Set & d [mm] 	& $<\!\Delta\phi_{1,2}\!>$ (FWHM) & $\mu$ 	      \\ \hline
 1 & 3.2	 	& $54.0^\circ$ 	      	    & $0.08\pm0.01$   \\
 2 & 4.5	 	& $38.1^\circ$ 	      	    & $0.23\pm0.02$   \\
 3 & 6.4	 	& $27.0^\circ$ 	      	    & $0.27\pm0.03$   \\
 4 & 7.2		& $24.1^\circ$ 	      	    & $0.29\pm0.04$   \\ \hline
 5 & 3.2 - 13.6		& $44.1^\circ$	      	    & $0.15\pm0.01$   \\
 6 & 4.5 - 13.6		& $31.0^\circ$	      	    & $0.25\pm0.01$   \\ 
 7 & 6.4 - 13.6 	& $23.7^\circ$	      	    & $0.27\pm0.02$   \\ \hline 
  \end{tabular}
  \caption{Modulation factor $\mu$ for $72^\circ\! \leq\! \theta_{1,2}\! \leq\! 90^\circ$, for different pixel distances $d$ and the corresponding mean azimuthal resolutions $<\!\Delta\phi_{1,2}\!>$. }
  \label{tab:mu_82}
\end{table}

The same analysis was repeated for scattering angles in range $60^\circ\! \leq\! \theta_{1,2}\! \leq\! 80^\circ$, centered around $\theta=70^\circ$, which was suggested as the optimal range with sufficient azimuthal correlation and abundant statistics \cite{toghyani2016}. This is summarized in Table \ref{tab:mu_70}.

\begin{table}
 \centering
 \begin{tabular}{c | c c c} \hline
  Set  &  d [mm]	& $<\!\Delta\phi_{1,2}\!>$ (FWHM)& $\mu$  \\ \hline
  1 & 3.2	 	& $54.0^\circ$ 	      	    & $0.05\pm0.01$   \\
  2 & 4.5	 	& $38.1^\circ$ 	      	    & $0.14\pm0.03$   \\
  3 & 6.4	 	& $27.0^\circ$ 	      	    & $0.26\pm0.03$   \\
  4 & 7.2		& $24.1^\circ$ 	      	    & $0.22\pm0.04$   \\ \hline
  5 & 3.2 - 13.6 	& $44.1^\circ$	      	    & $0.12\pm0.01$   \\
  6 & 4.5 - 13.6 	& $31.0^\circ$	      	    & $0.17\pm0.01$   \\
  7 & 6.4 - 13.6 	& $23.7^\circ$	      	    & $0.21\pm0.02$   \\ \hline
   
  \end{tabular}
  \caption{Modulation factor $\mu$ for $60^\circ\! \leq\! \theta_{1,2}\! \leq\! 80^\circ$, for different pixel distances $d$ and the corresponding mean azimuthal resolutions $<\!\Delta\phi_{1,2}\!>$.}
  \label{tab:mu_70}
\end{table}

The results show the modulation of the $\phi_1-\phi_2$ distribution, as expected due to initial orthogonality of polarizations of the annihilation quanta. The strength of the modulation depends on the scattering angles $\theta_{1,2}$ and we have indeed observed a stronger modulation for scattering around $\theta_{1,2}=82^\circ$, than for scattering around $\theta_{1,2}=70^\circ$. The modulation also depends on the angular resolution and we have observed stronger modulation for smaller $\Delta\phi_{1,2}$. Such behaviour is expected, since the finite geometries reduce the effective modulation strength with respect to the one that would be obtained for an infinite precision in $(\theta,\phi)$\cite{snyder1948}. 

The observed ratio of identified Compton events to events with full-energy deposition in single pixel is $R_{CE}=1.6\%$ for $72^\circ\! \leq\! \theta_{1,2} \!\leq\! 90^\circ$, and $R_{CE}=2.6\%$ for $60^\circ\! \leq\! \theta_{1,2} \!\leq\! 80^\circ$. These ratios are modest, which should not come as a surprise, since LFS is optimized for high photo-electric cross section and a high stopping power, that enables standard PET devices to dominantly exploit the abundant single-pixel events.

In order to increase the polarimetric sensitivity, a better angular resolution should be provided. The improvement in $\Delta\theta$ may be achieved by improving the energy resolution, while the one in $\Delta\phi$ could be achieved either by finer segmentation or by using a detector material with lower stopping power that would allow more Compton events with more distant pixels fired. A promising candidate is GAGG:Ce (e.g. \cite{shimazoe2018, uenomachi2018}), which offers a superior energy resolution, as well as lower density and lower effective atomic number than LFS or LYSO, which should also result in a larger $R_{CE}$, desirable in this concept. 

\section{Conclusions}
We have used a system of two compact, position and energy-sensitive, single-layer Compton detectors to investigate the feasibility of measuring the polarization correlations of annihilation quanta. The coincidence data from positron annihilations has been collected and events with Compton scattering in both modules are selected. The polarimetric modulation has been extracted from the difference in the azimuthal scattering angles of the two gammas, demonstrating the feasibility of the approach. Although a moderate polarimetric sensitivity has been observed, it may be improved by optimizing detector material and geometry to provide better angular resolutions. Such detectors might be exploited in PET or other experiments where measurement of gamma polarization is of interest. Importantly, the detectors based on the single-layer concept would significantly improve the cost-efficiency compared to typical two layer systems used for Compton scattering detection.

\section*{Acknowledgements}
This work has been supported in part by Croatian Agency for Small and Medium Enterprises, Innovations and Investments (HAMAG-BICRO) project POC6\_1\_211, in part by Croatian Science Foundation under the project 8570 and in part by International Atomic Energy Agency under the project CRP F22069. 

\bibliographystyle{unsrt}
\bibliography{references_short}

\begin{thebibliography}{10}

\bibitem{lei1997}
F.~Lei et~al.
\newblock Compton polarimetry in gamma-ray astronomy.
\newblock {\em Space Sci. Rev.}, 82:309--388, 1997.

\bibitem{mitani2004}
T.~Mitani et~al.
\newblock {A prototype Si/CdTe Compton camera and the polarization
  measurement}.
\newblock {\em IEEE Trans. Nucl. Sci.}, 51(5):2432--2437, 2004.

\bibitem{spillmann2008}
U~Spillmann et~al.
\newblock {Performance of a Ge-microstrip imaging detector and polarimeter}.
\newblock {\em Rev. Sci. Instr.}, 79(8):083101, 2008.

\bibitem{bloser2009}
P.~F. Bloser et~al.
\newblock {{Calibration of the Gamma-RAy Polarimeter Experiment (GRAPE) at a
  Polarized Hard X-Ray Beam}}.
\newblock {\em Nucl. Instrum. Methods Phys. Res. A}, 600:424--433, 2009.

\bibitem{takeda2010}
S.~Takeda et~al.
\newblock Polarimetric performance of {Si/CdTe} semiconductor {Compton} camera.
\newblock {\em Nucl.Instrum.Methods Phys.Res.A}, 622:619--627, 2010.

\bibitem{yonetoku2011}
D.~Yonetoku et~al.
\newblock {Gamma-Ray Burst Polarimeter - GAP - aboard the Small Solar Power
  Sail Demonstrator IKAROS}.
\newblock {\em Publ. Astron. Soc. Jap.}, 63:625--638, 2011.

\bibitem{kuncic2011}
Zdenka Kuncic et~al.
\newblock {Polarization enhanced X-ray imaging for biomedicine}.
\newblock {\em Nucl. Instrum. Methods Phys. Res. A}, 648:S208--S210, 2011.

\bibitem{mcnamara2014}
Aimee McNamara et~al.
\newblock {Towards optimal imaging with PET: An in silico feasibility study}.
\newblock {\em Phys. Med. Biol.}, 59:7587--7600, 11 2014.

\bibitem{toghyani2016}
M~Toghyani et~al.
\newblock {Polarisation-based coincidence event discrimination: An in silico
  study towards a feasible scheme for Compton-PET}.
\newblock {\em Phys. Med. and Biol.}, 61:5803--5817, 08 2016.

\bibitem{pryce1947}
M.~H.~L. Pryce and J.~C. Ward.
\newblock {{Angular Correlation Effects with Annihilation Radiation}}.
\newblock {\em Nature}, 160:435, 1947.

\bibitem{snyder1948}
H.~S. Snyder, S.~Pasternack, and J.~Hornbostel.
\newblock {{Angular Correlation of Scattered Annihilation Radiation}}.
\newblock {\em Physical Review}, 73:440, 1948.

\bibitem{makek2017}
M.~Makek et~al.
\newblock {Performance of scintillation pixel detectors with MPPC read-out and
  digital signal processing}.
\newblock {\em Acta Physica Polonica B}, 48(10):1721--1726, 2017.

\bibitem{makek2019}
M.~Makek, D.~Bosnar, and L.~Paveli\'{c}.
\newblock {Scintillator Pixel Detectors for Measurement of Compton Scattering}.
\newblock {\em Condensed Matter}, 4(1):24, 2019.

\bibitem{shimazoe2018}
K.~Shimazoe et~al.
\newblock Development of simultaneous {PET} and {Compton} [imaging using
  {GAGG-SiPM} based pixel detectors.
\newblock {\em Nucl. Instrum. Methods Phys. Res. A}, 2018.

\bibitem{uenomachi2018}
M.~Uenomachi et~al.
\newblock {Double photon emission coincidence imaging with GAGG-SiPM Compton
  camera}.
\newblock {\em Nucl. Instrum. Methods Phys. Res. A}, 2018.

\end{thebibliography}

\end{document}